\documentstyle[aps,prc,epsf,amssymb]{revtex}

\begin{document}
\draft
\def\vec#1{{\bf{#1}}}
\bibliographystyle{srt}

\wideabs{
\title{
		Investigation of the Exclusive
		$^{3}{\text{He}}(e,e^{\prime}pp)n$ Reaction}

\author{
	D.L.~Groep$^{1}$, 
	M.F. van~Batenburg$^{1}$, 
	Th.S.~Bauer$^{1,3}$, 
	H.P.~Blok$^{1,2}$, 
	D.J.~Boersma$^{1,3}$, 
	E.~Cisbani$^{4}$, 
	R.~De Leo$^{5}$, 
	S.~Frullani$^{4}$, 
	F.~Garibaldi$^{4}$, 
	W.~Gl\"ockle$^{6}$, 
	J.~Golak$^{7}$, 
	P.~Heimberg$^{1,2}$, 
	W.H.A.~Hesselink$^{1,2}$, 
	M.~Iodice$^{4}$, 
	D.G.~Ireland$^{8}$, 
	E.~Jans$^{1,}$\cite{co}, 
	H.~Kamada$^{6}$, 
	L.~Lapik\'as$^{1}$, 
	G.J.~Lolos$^{9}$, 
	C.J.G.~Onderwater$^{1,2,}$\cite{goaddr},
	R.~Perrino$^{10}$, 
	A.~Scott$^{8}$, 
	R.~Starink$^{1,2}$, 
	M.F.M.~Steenbakkers$^{1,2}$, 
	G.M.~Urciuoli$^{4}$, 
	H. de~Vries$^{1}$, 
	L.B.~Weinstein$^{11}$, 
	H.~Wita{\l}a$^{7}$
}

\address{$^{1}$NIKHEF, P.O.~Box~41882, 1009~DB~Amsterdam, The~Netherlands}
\address{$^{2}$Vrije~Universiteit, de~Boelelaan 1081, 1081~HV~Amsterdam,
	The~Netherlands}
\address{$^{3}$Universiteit~Utrecht, P.O.~Box~80.000, 3508~TA~Utrecht,
	The~Netherlands}
\address{$^{4}$Istituto~Superiore~di\
	Sanita\makebox[0pt]{\hspace*{3pt}\`~}, Laboratorio~di~Fisica, INFN,
	Viale~Regina~Elena~299, Rome, Italy}
\address{$^{5}$INFN~Sezione~di~Bari, Dipartimento~Interateneo\
	di~Fisica, Via~Amendola~173, Bari, Italy}
\address{$^{6}$Institut~f\"ur~Theoretische~Physik~II,
	Ruhr-Universit\"at~Bochum, D-44780~Bochum, Germany}
\address{$^{7}$Institute of Physics, Jagellonian University, 
	PL-30059 Cracow, Poland}
\address{$^{8}$Department~of~Physics~and~Astronomy, University\
	of~Glasgow, Glasgow G12 8QQ, UK}
\address{$^{9}$Department~of~Physics, University~of~Regina, Regina
	SK~S4S~0A2, Canada}
\address{$^{10}$INFN~Sezione~di~Lecce, via per Arnesano, 73100 Lecce, Italy}
\address{$^{11}$Physics~Department, Old~Dominion~University, Norfolk,
	Virginia~23529}

\date{\today}

\maketitle

\begin{abstract}
	Cross sections for the $^{\text{3}}{\text{He}} (e,e^{\prime}pp)n$
	reaction were measured over a wide range of energy and
	three-momentum transfer.  At a momentum transfer $q$=375~MeV/$c$,
	data were taken at transferred energies $\omega$ ranging from 170
	to 290~MeV.  At $\omega$=220~MeV, measurements were performed at
	three $q$ values (305, 375, and 445~MeV/$c$).  The results are
	presented as a function of the neutron momentum in the
	final-state, as a function of the energy and momentum transfer,
	and as a function of the relative momentum of the two-proton
	system.  The data at neutron momenta below 100~MeV/$c$, obtained
	for two values of the momentum transfer at $\omega$=220~MeV, are
	well described by the results of continuum-Faddeev calculations. 
	These calculations indicate that the cross section in this domain is
	dominated by direct two-proton emission induced by a one-body
	hadronic current.  Cross section distributions determined as a
	function of the relative momentum of the two protons are fairly
	well reproduced by continuum-Faddeev calculations based on various
	realistic nucleon-nucleon potential models.  At higher neutron
	momentum and at higher energy transfer, deviations between data and
	calculations are observed that may be due to contributions of
	isobar currents.
\end{abstract}
%
%
} 

\section{Introduction}

The study of exclusive two-nucleon emission by electrons at
intermediate energies provides a tool to investigate the role of
nucleon-nucleon correlations inside atomic nuclei.  Advances in the
theoretical description of light nuclei and of few-nucleon reaction
processes based on modern nuclear forces allow for detailed
comparisons of experimental results with
calculations~\cite{glo96,car98}.  In few-nucleon systems, the
Schr\"odinger equation, expressed in the form of Faddeev-Yakubovsky
equations, can be solved exactly and the calculations can be performed
based on realistic $NN$ interactions like Bonn-B, CD Bonn, Argonne
$v_{18}$ and Nijmegen-93, I and II.  The agreement between theory and
data for binding energies~\cite{Nogga} and low energy spectra, as well
as three-nucleon scattering observables, is in most cases quite
remarkable. 

A recent review of applications of Faddeev equations for the
three-nucleon bound and scattering states occurring in processes
induced by electromagnetic probes and based on realistic forces can be
found in Ref.~\cite{glo99}.  Here we report on a measurement of the
$^{3}\mbox{He}(e,e^{\prime}pp)n$ reaction performed at the AmPS
facility at NIKHEF.  Data will be compared to theoretical results
achieved by the Bochum-Cracow collaboration.

The cross section for electron-induced two-nucleon knockout at
intermediate electron energies is driven by several processes.  The 
\emph{NN} interaction at small inter-nucleon distances induces strong
correlations between the nucleons inside
the nucleus, which influences the momentum distributions of the bound
nucleons and consequently the knock-out of nucleons by the
absorption of a virtual photon via a one-body electromagnetic current. 
The interaction of the virtual photon with two-body currents, either
via coupling to mesons or via intermediate $\Delta$ excitation, will
also contribute to the cross section for one- and two-nucleon
knockout. In the latter case, this contribution is expected
to be largest if the virtual photon couples to a proton-neutron pair. 
In addition, final-state interactions\ (FSI) among the ejectiles
influence the two-nucleon knockout cross section. 

The availability of exact calculations and the well-defined final
state make the tri-nucleon system a good candidate for two-nucleon
knockout studies, since one may attempt to unravel the tightly
connected ingredients of the reaction, especially short-range
correlations, two-body currents and FSI by comparison of data,
measured under various kinematic conditions, with the results of these
calculations. 

In order to investigate the reaction mechanism of two-nucleon emission
and the relative importance of one-body and two-body hadronic currents
in the cross section, measurements should be performed over a
wide range of energy-transfer values covering the domain from the
`dip' region to the $\Delta$-resonance.  In addition, the coupling
mechanism of the virtual photon to the tri-nucleon system may be
investigated with measurements performed at various values of the
three-momentum transfer. 

In order to extract information on the relative and center-of-mass
motion of the nucleon pair, measurements covering a large angular
domain and a sufficient range in kinetic energy have to be performed
in a kinematic domain in which the contribution of one-body hadronic
currents to the cross section dominates.

Two-proton emission from $^{3}\mbox{He}$ has been studied before using
photons produced via
bremsstrahlung. The measurements were performed by Audit
\emph{et~al.}~\cite{aud89,aud91} in a kinematic domain selected to
emphasize the production of on-shell pions on the struck nucleon that
are subsequently reabsorbed on the nucleon pair.  The results were
evaluated in a theoretical framework based on a diagrammatic expansion
of the reaction amplitude~\cite{lag85}.  These measurements indicated
an important role for processes in which three nucleons are involved,
in particular, a sequential pion exchange.  Such processes were also
observed at lower energy transfers, in which the initial pion is
assumed to propagate off-shell~\cite{sar93}.

The use of tagged photon beams opened the possibility to perform
kinematically complete measurements of the cross section for full
breakup.  Data obtained with the large-solid-angle detector
DAPHNE~\cite{aud97} in the $\Delta$ resonance region, showed that the
cross section for photon-induced breakup at $E_\gamma<500~\text{MeV}$
is dominated by two-step three-nucleon processes in those regions of
phase space where final-state rescattering effects are minimal.  No
neutron momentum distribution could be extracted from this data set. 
The role of three-nucleon mechanisms was also observed by Kolb
\emph{et al.}~\cite{kol96} at lower photon energies.  Neutron momentum
distributions extracted from the
$^3\mbox{He}(\gamma,pp)n$~\cite{emu94} and
$^3\mbox{He}(\gamma,pn)p$~\cite{emu94b} reactions by Emura \emph{et
al.} in the $\Delta$-resonance region~($E_\gamma$=200--500~MeV and
125--425~MeV, respectively) showed that both two-nucleon and
three-nucleon photo-absorption mechanisms are needed to explain the
data, but that at low neutron momentum the two-nucleon processes
dominate the cross section.  The choice of the kinematic domain and
the transverse nature of the probe used in these experiments caused
that the absorption of the photon by a two-proton pair was found to be
largely driven by two-body hadronic currents.

The study of $NN$ correlations by means of the $(e,e'pp)$ reaction was
pioneered at NIKHEF in the $^{12}\text{C}(e,e'pp)$ experiments by
Zondervan \emph{et al.}~\cite{zon95} and Kester \emph{et
al.}~\cite{kes95}.  The advance of high duty-cycle electron
accelerators has made possible the three-fold coincidence experiments
necessary to measure exclusive electron-induced two-nucleon knockout. 
Measurements performed by Onderwater~\emph{et
al.}~\cite{ond97,ond98art} at the Amsterdam Pulse Stretcher facility
AmPS using large-solid-angle proton detectors, revealed clear
signatures of short-range correlations in the
$^{16}\text{O}(e,e'pp)^{14}\text{C}$ reaction.  Similar results were
obtained with a three-spectrometer setup at the Mainz Microtron
MAMI~\cite{ros97}.  Experimental evidence for short-range correlations
was obtained by Starink \emph{et al.}~\cite{sta99art} from the
measured energy-transfer dependence of the
$^{16}\text{O}(e,e'pp)^{14}\text{C}_{\mathrm{g.s.}}$ reaction.

In this paper, we present results of a $^{3}\text{He}(e,e^{\prime}pp)n$
experiment performed at various values of the four-momentum transfer
$(\omega,q)$ of the virtual photon.  Energy and three-momentum
transfer of the virtual photon were varied over a kinematic domain
ranging from the `dip' region to just below the $\Delta$ resonance.
A partial account of the present work, which is the first study of
electron-induced two-proton knockout out $^{3}\mbox{He}$, has been
published in Ref.~\cite{groprl}, where only the data taken at
$\omega$=220~MeV were discussed.

\section{Theoretical framework}
\label{sec:theory}

In the exclusive electron-induced two-nucleon knockout reaction,
energy and momentum are transferred to a nucleus by a virtual photon,
while after the reaction the momenta of the scattered electron and
both nucleons are determined.  Here we consider only those processes
where the remainder of the nucleus is left intact and no secondary
particles are created.

\subsection{Kinematics and Reaction Mechanism}

\begin{figure}
    \centerline{\epsffile{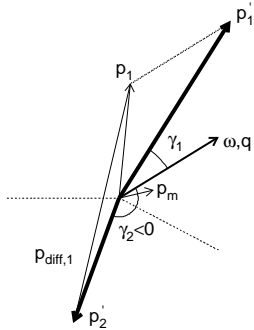}}
    \caption{
		Kinematic configuration of an $(e,e'pp)$ reaction. For
		clarity all momentum vectors are shown in one plane. Dotted
		lines represent the incoming and outgoing electron, the bold 
		vectors the detected protons, and the thin vectors derived 
		quantities. The reconstructed momentum vectors $\vec{p}_1$ and
		$\vec{p}_{\mathrm{diff,1}}$ assume coupling of the virtual
		photon to proton-1.
		\label{fig:kinema}
	}
\end{figure}

The kinematics for the $^{3}\text{He}(e,e^{\prime}pp)n$
reaction is schematically shown in Fig.~\ref{fig:kinema}.  Within the
one-photon exchange approximation the exchanged virtual photon
carries an energy $\omega=E_e-E_{e'}$ and a three-momentum
$\vec{q}=\vec{p}_e-\vec{p}_{e'}$.  The two protons, with momenta
$\vec{p}'_1$ and $\vec{p}'_2$, emitted after the full breakup of
$^{3}\mbox{He}$, are detected in coincidence with the scattered
electron.  Proton--1 is emitted at the smallest
angle $\gamma_1$ with respect to $\vec{q}$ and is referred to as the
`forward' proton.  The second proton, emitted opposite to
$\vec{q}$ with an angle $\gamma_2$ is referred to as the `backward'
proton. 

In an exclusive $^{3}\mbox{He}(e,e'pp)n$ experiment, the final state can be
reconstructed completely and the missing momentum
\begin{equation}
	\vec{p}_m=\vec{q}-\vec{p}'_1-\vec{p}'_2
\end{equation}
is equal to the momentum of the undetected neutron~$\vec{p}'_3$.  Energy
conservation requires that the missing energy
\begin{equation}
E_m=\omega-T_1-T_2-T_{\mathrm{rec}}
\end{equation}
be equal to the binding energy $E_b$ of the $^{3}\mbox{He}$ nucleus;
the excitation energy $E_x=0$ if the reaction is confined to
two-nucleon emission.  Here, $T_1$ and $T_2$ are the
kinetic energies of the two emitted protons, and $T_{\mathrm{rec}}$ is
the kinetic energy of the recoiling neutron, which can be calculated
from $\vec{p}_m$.

There are various ways the virtual photon can couple to the
$^{3}\mbox{He}$\ nucleus.  The one-body hadronic current accounts for
the absorption of a virtual photon by one nucleon of a correlated
pair, which subsequently leads to the full breakup of the tri-nucleon
system.  If one assumes the coupling of the virtual photon to the
proton emitted in forward direction (proton--1), one can define the
momentum difference $p_{\mathrm{diff,1}}= |\vec{p}'_1-\vec{q}-\vec{p}'_2|$,
which in this case can be identified with $2 (\vec{p}_1-\vec{p}_2)$,
where $\vec{p}_1$ and $\vec{p}_2$ are the momenta of the protons in
the initial state.

Breakup of the $^{3}\mbox{He}$ nucleus can also occur via two-body
hadronic currents, thus sharing the transferred momentum
between two nucleons.  In the energy- and momentum-transfer domain
under study, two-body currents are
involved in meson-exchange (MEC) and excitation of the
$\Delta$ resonance followed by the decay $\Delta N \rightarrow NN$
(isobar currents or ICs).  Their importance strongly depends on the
isospin of the $NN$ pair.  In the case of a $pp$ pair, the contribution
of MECs to the cross section will be strongly suppressed, as the
virtual photon, in a non-relativistic reduction of the current
operator, does not couple to such a
pair~\cite{giu91}.  Also the contribution due to isobar currents is
reduced for two protons in a relative $^1S_0$ state, as the transition
via the resonant M1 multipole is forbidden by angular-momentum and
parity-conservation rules.  Therefore $\Delta$-excitation is only
possible via the much weaker C2 and E2 multipoles~\cite{wil96}.  These
restrictions on MECs and ICs do not apply to $pn$ pairs.  It may
therefore be expected that in a direct $(e,e'pp)$ reaction the
influence of these two-body currents is reduced compared to the
$(e,e'pn)$ case.

The photon can also couple to all three particles by a
three-body mechanism.  Sensitivity to these processes will exist at
photon energies around 500--600~MeV and in specific regions of phase
space, where the struck meson initially propagates on-shell and is
subsequently reabsorbed by the remaining nucleon pair~\cite{bof96}.

\subsection{Calculations}

The differential cross section of the electron-induced three-body
breakup of the tri-nucleon system has been calculated by solving
consistently Faddeev-type equations for both the bound state and the
final `scattering' state using the same nuclear
forces~\cite{glo96,mei86}.  Both theoretical~\cite{mei86,ish94} as
well as experimental studies~\cite{poo99,ank98,spa98} indicate that
rescattering among the outgoing nucleons has a significant effect on
the cross sections and the spin asymmetries.  The measured cross
sections are compared to the results of these `continuum Faddeev'
calculations, which completely account for these rescattering effects
in the final state~\cite{gol95}.  The calculations are based on the
potential models Bonn-B, CD-Bonn, Argonne $v_{18}$ and Nijmegen-93,
and were performed with both a non-relativistic one-nucleon current
operator and with this one-body current augmented by two-body current
operators. 

From the two-body hadronic currents, only meson-exchange currents have
been incorporated, using a formalism like in~\cite{sch89}, which
includes coupling to one-pion and one-$\rho$ exchange.  In order to
incorporate these currents in a way compatible with the potential
model used, the prescription of~\cite{ris85} is used.  For technical
details see~\cite{kot99}.

At present, first attempts to account for $\Delta$-excitation and
deexcitation within the continuum Faddeev framework are
available~\cite{chm99}, but up to now applications to electromagnetic
processes have not been published.  However, ample evidence exists
that isobar currents are the dominant two-nucleon knockout mechanism
at values of the $\gamma^{*}NN$ invariant mass around that of the
$N\Delta$ system, and are responsible for a large part of the direct
$pn$ emission strength also at lower energy-transfer
values~\cite{mac93}.  Discussion of isobar contributions to the
measured cross sections for $^{3}\mbox{He}(e,e'pp)n$ is deferred to
section~\ref{sec:res}.

\begin{figure}
    \centerline{\epsffile{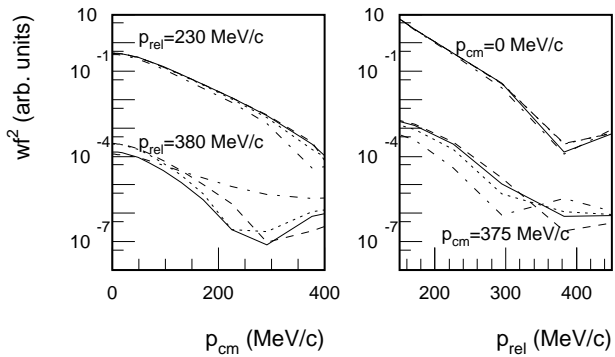}}
    \caption{
		Probability distribution of the $^{3}$He bound-state wave
		function for a two-nucleon pair in a relative $^1S_0$ state,
		on the left-hand side displayed as a function of the
		center-of-mass momentum and on the right-hand side 
		as a function of the relative
		momentum in the nucleon pair.  The curves show calculations
		with the Bonn-B (solid), CD-Bonn (dashed), Nij\-me\-gen-93
		(dotted) and Argonne $v_{18}$ (dot-dashed) potential
		models. The domain shown is indicative for the region of phase
		space covered in this experiment.
		\label{fig:wavefunc}
	}
\end{figure}

The bound-state wave function of $^{3}\mbox{He}$ has been calculated
using Faddeev techniques, based on realistic models of the interaction
between two nucleons~\cite{nog97}.  In the calculations for the
bound-state wavefunction, shown in Fig.~\ref{fig:wavefunc}, 34
channels were considered.  Various types of calculations were
performed, based on different models of the $NN$-interaction.  Results
are shown for the Bonn-B, charge dependent~(CD) Bonn, Argonne $v_{18}$
and Nijmegen-93 potential models.  Large differences are observed
between the various calculations both at high relative
($p_{\mathrm{rel}}$) and at high center-of-mass ($p_{\mathrm{cm}}$)
momenta of the nucleons. 

Direct information on the initial $^{3}\mbox{He}$ bound state can in
principle be obtained from reactions induced by a one-body hadronic
current; in this case, the momentum of the virtual photon is
transferred to a single nucleon only.  In absence of final-state
rescattering, this implies that the non-struck particles have equal
momenta in the final and the initial state, and that the
initial-state configuration can be reconstructed exactly from the
measured proton momenta.  However, the knowledge of which nucleon was
hit cannot be deduced from the data, as the measured cross section
(even in the plane-wave impulse approximation, PWIA) is the coherent
sum of the transition amplitudes describing coupling of the virtual
photon to any of the three nucleons.

Nevertheless, insight into the coupling mechanism can been deduced from
the PWIA calculations.  Although the cross sections calculated with this
model cannot be compared to experimental data, they are valuable to
determine the relative importance of the coupling of the virtual
photon to the various particles leading to the same final state.  A
comparison of the cross sections, calculated assuming coupling to only
one nucleon, shows that, over the entire energy acceptance, the
reaction is dominated by coupling to the forward proton.

\section{Experiment and Analysis}

The measurements were performed with the electron beam provided by the
Amsterdam Pulse Stretcher facility~(AmPS), with a macroscopic duty
factor of 70--80\%.  The incident electron energy, as determined from
elastic scattering experiments, was 564~MeV and the beam current
varied between 0.5 and 1.5~$\mu\mbox{A}$, depending on the kinematic
setting.  The target setup consisted of a cryogenic, high-pressure
`barrel' cell containing gaseous $^{3}\mbox{He}$ with a nominal
thickness of 268~$\mbox{mg}\,\mbox{cm}^{-2}$.  A graphite target and
an aluminium-oxide target were used for beam-calibration purposes and
detector efficiency determination.

The scattered electrons were detected in the QDQ magnetic
spectrometer.  This focussing spectrometer can detect electrons within
a range of $\pm$4.5\% with respect to the selected central momentum
value.  Its momentum resolution is better than $2\times
10^{-4}$~\cite{vri90}.  The solid angle is defined by an octangular
slit with an acceptance of $\pm$70~mrad in both the in-plane as well
as out-of-plane direction. 

To detect the protons emitted from the target two scintillator
detectors were used:  HADRON3 and HADRON4.  The design of both HADRON
detectors is similar.  They both cover a large solid angle (230 and
540~msr, respectively) and span a sizeable range in detected proton
energies. These ranges are 72 to 255~MeV for HADRON3, shielded by 
5.2~mm of lead, and 47 to 180~MeV for HADRON4, shielded by 2.0~mm 
of stainless steel~\cite{pel99}. 
The fine-grain segmented layout, that is necessary to limit the
counting rate in the individual elements to less than 1~MHz, 
provides an angular resolution of 0.5~(1)~degrees in-plane and
1~(2)~degree out-of-plane for HADRON3~(HADRON4). The proton-energy
resolution amounts to 2.5\% (FWHM).

In an $(e,e'pp)$ reaction, each of the three detectors~(QDQ, HADRON3
and HADRON4) generates a trigger upon arrival of a particle meeting
the detector-specific requirements.  To determine the number of real
and accidental events within the three-fold coincidence time region, a
\emph{coincidence detector} is employed, which measures the arrival
times of the three detector trigger signals with a resolution of
49~ps.  Each trigger starts a time window with a length of 125~ns. 
The latter signals are used by the gating and prescaling module to
classify events based on their coincidence type~(single, double or
triple).  The read-out of the various types of events is completely
independent as long as the total event rate does not exceed 5~kHz or
1.4~MByte/s.  In this experiment all three-fold coincident
events~(triples) were stored as well as a subset of the doubles and
singles for monitoring purposes.

\subsection{Kinematic conditions}

In order to enhance the contribution to the cross section due to
knockout of correlated proton pairs, measurements were performed in
the so-called `dip' region.  Here, the knocked-out protons receive
sufficient energy to pass the threshold for detection in both HADRON
detectors, while at higher energy transfers the contribution from
$\Delta$ excitation is expected to increase.

Measurements were performed at various values of the three-momentum
transfer $q$, to investigate the coupling mechanism of the virtual
photon to the $^{3}\mbox{He}$ system.  The values of transferred
momentum were:  305~(LQ), 375~(CQW) and 445~(HQ)~MeV/$c$, where the
choice of the maximum value of 445~MeV/$c$ was dictated by the
strongly reduced count rate at higher values of $q$.  In these
measurements $\omega$ was kept constant at 220~MeV.

A series of measurements at various values of the energy transfer,
ranging from 170 to 290~MeV, allowed the study of the reaction mechanism
as a function of the invariant energy of the photon and two-proton
final state.  In this way the relative importance of one-body and
two-body hadronic currents was investigated.  The measurements were
performed at $q$=375~MeV/$c$.  Within the range 170--290~MeV the
invariant mass of the two-proton system in the final state,
$W_{p_1^{\prime}{p_2^{\prime}}}$, varies from 2005 to 2120~MeV/$c^2$. 

The forward proton detector, HADRON3, was positioned such that the
angle between $\vec{q}$ and $\vec{p}_1$ was minimal, within the
geometrical restrictions of detector housing and beam pipe.  The
positioning of the second proton detector was guided by the kinematics
of quasi-free two-proton emission, where at $p_m=0~\mbox{MeV}/c$ the
protons are emitted at conjugate angles. 

An overview of all employed kinematic settings is given in
Table~\ref{tab:kinema}.  At the LQ kinematic setting, which features
the largest flux of virtual photons, additional measurements at other
proton angles were performed to investigate the angular correlation
and the behaviour of the cross section as a function of $\gamma_1$,
the angle between $\vec{q}$ and the forward proton.

\subsection{Analysis}

\begin{figure}
    \centerline{\epsffile{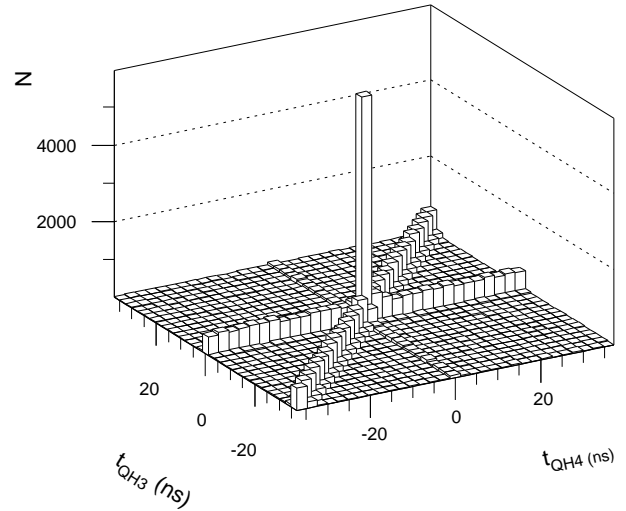}}
    \caption{
		Time difference distribution for three-fold coincident
		events, as measured in kinematics LQA ($\omega$=220~MeV, 
		$q$=305~MeV/$c$).
		\label{fig:cptlqa}
	}
\end{figure}

Using established procedures, the momenta of the scattered
electron~\cite{vri90} and of both emitted protons~\cite{pel99} were
determined.  For each of the three detectors, the trigger arrival time
at the coincidence detector is measured and corrected off-line for
time-of-flight differences and detector-specific delays. 
Based on the arrival-time differences, as shown in
Fig.~\ref{fig:cptlqa}, various types of events can be distinguished; 
the peak, located at a time difference of 0~ns for both QDQ--HADRON3
%
%
\begin{table}
	\caption{
		Overview of the kinematic configurations of the
		$^{3}\mbox{He}(e,e'pp)$~experiment.  The incident energy was
		563.7~MeV. 
		\label{tab:kinema}
	}

\begin{center}
\begin{tabular}{lccrrr}
label	&
	\multicolumn{1}{p{0.41in}}{\centering$\omega$\\(MeV)} & 
	\multicolumn{1}{p{0.52in}}{\centering$q$\\(MeV/$c$)} & 
	\multicolumn{1}{p{0.37in}}{\centering$\theta_{e^\prime}$\\(deg)} & 
	\multicolumn{1}{p{0.31in}}{\centering$\theta_{\text{H3}}$\\(deg)} & 
	\multicolumn{1}{p{0.40in}}{\centering$\theta_{\text{H4}}$\\(deg)} \\
\hline
LQA	&	220	&	305	& $-27.72$	&	53.8	&	$-120.4$	\\
LQV	&	220	&	305	& $-27.72$	&	53.8	&	$-92.9$		\\
PEF	&	220	&	305	& $-27.72$	&	79.9	&	$-100.1$	\\
CQW	&	220	&	375	& $-40.26$	&	53.8	&	$-105.3$	\\
HQ	&	220	&	445	& $-52.01$	&	53.8	&	$-119.7$	\\
LW	&	190	&	375	& $-41.14$	&	53.8	&	$-119.7$	\\
IW	&	250	&	375	& $-38.72$	&	53.8	&	$-105.3$	\\
HW	&	275	&	375	& $-36.76$	&	53.8	&	$-105.3$	\\
\end{tabular}
\end{center}
\end{table}
\noindent
and QDQ--HADRON4 coincidences, corresponds to real three-fold coincidences, 
is superimposed on a background due to double coincident and single
events.  The two ridges at $\Delta t_{\text{QH4}}$=0~ns and $\Delta
t_{\text{QH3}}$=0~ns are due to real $(e,e'p)$ coincidences between
the scattered electron and either the backward or forward proton
detector together with an accidental second proton.  The ridge at
$\Delta t_{\text{QH4}}=\Delta t_{\text{QH3}}$ contains real two-proton
coincidences together with an accidental electron trigger.  The
structures sit on a flat background of events that are
three-fold uncorrelated.

To extract the true $(e,e'pp)$ events, the contributions
of the flat background and the ridges to the region of the real
coincidences has to be estimated. This is best performed by
symmetrizing the coincidence time spectrum by a linear
transformation
\begin{eqnarray}
  \tau_x & = & \frac{1}{\sqrt{3}} \,\, \bigl( 2 t_{\text{H4}} - t_{\text{Q}}
    - t_{\text{H3}} \bigr) \\
  \tau_y & = & \bigl( t_{\text{Q}} - t_{\text{H3}} \bigr).
\end{eqnarray}
After this transformation, the time difference distribution will
exhibit a symmetric hexagonal shape as displayed in
Fig.~\ref{fig:karrewiel}.  The width $b$ is chosen such that the full
widths of the two-fold coincidence bands are well within the regions $B$. 
As the coincidence time resolution is always better than
1.5~ns~(FWHM), $b$ was chosen to be 3~ns, \emph{i.e.}, at least
4.5$\sigma$ from the peak position.  

The number of true $(e,e'pp)$ events is obtained by subtracting the
number of accidental coincidences determined from the regions $B$ and
$C$, from the number of events in the $A$ region:
\begin{figure}
    \epsfxsize=2.375in
    \centerline{\epsffile{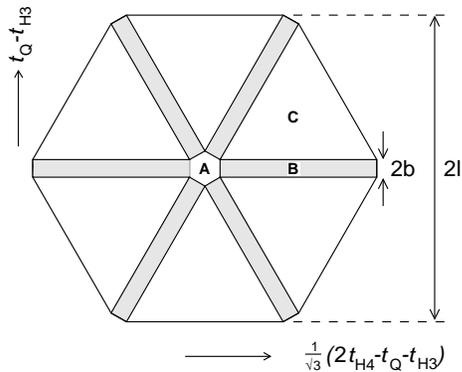}}
    \caption{
		A hexagonal shape is obtained in the $(\tau_x,\tau_y)$
		plane, after the coincidence time distributions are symmetrized.
		\label{fig:karrewiel}
	}
\end{figure}
\begin{equation}
        N_T = N_A - f_B N_B - f_C N_C,
\end{equation}
where the fractions $f_B$ and $f_C$ are derived from the relative
lengths and surfaces of the regions $B$ and $C$ with respect to $A$. 
Events outside the regions $A$, $B$ and $C$ are discarded.  The value
of $l$ determines the accuracy with which the number of accidental
coincidences in the region $A$, where the real coincidences are
located, can be estimated.  This length $l$ was chosen to be 60~ns. 

\begin{figure}
    \centerline{\epsffile{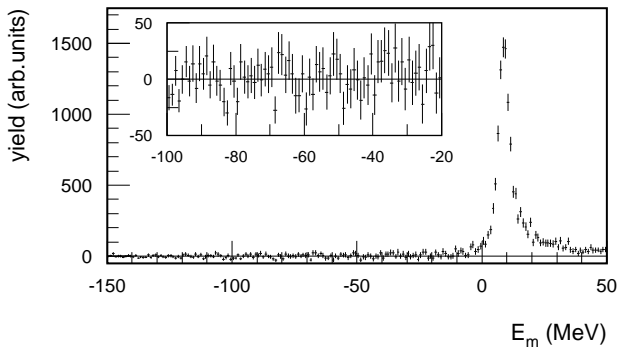}}
    \caption{
		Missing-energy distribution for kinematics LQA. The
		resolution amounts to 5.5~MeV (FWHM). The inset shows an
		enlargement of a subset of the same distribution.
		\label{fig:emlqa}
	}
\end{figure}

The method used to subtract the accidental coincidences can be
verified by inspecting the missing-energy distribution of the resulting
$(e,e'pp)$ coincidences.  Below the value corresponding to the
two-proton separation energy, in this case the binding energy of
$^{3}\mbox{He}$, no true $(e,e'pp)$ events can occur.  The
missing-energy distribution of true $(e,e'pp)$ events in kinematics
LQA is shown in Fig.~\ref{fig:emlqa}.  The peak corresponding to the
three-body breakup of $^{3}\mbox{He}$ is located at $E_m$=7.7~MeV. 
The inset shows an enlargement of the $E_m$ distribution
for the range $-100\!<\!E_m\!<\!-20~\text{MeV}$.  The yield in this
region, -1.1$\pm$1.7, is consistent with zero.

The eight-fold differential cross section for the reaction
$^{3}\mbox{He}(e,e'pp)$ is determined as a function of various
kinematic quantities like $\omega$ or $p_m$.  It is calculated as
\begin{equation}
  \frac{d^8\sigma}{dV^8}(\Delta\vec{X}) = 
    \int_{E_x} \frac{N(\Delta\vec{X})} 
    {\int\!\!{\cal{L}}\,dt \: {\cal V}(\Delta\vec{X}) }
      \left| \frac{\partial T_2}{\partial E_x} \right| dE_x.
  \label{eq:xsformal}
\end{equation}
In this equation, $\Delta\vec{X}$ refers to a range of values of (a
set of) kinematic quantities in which the cross section will be
represented, \emph{e.g.}, $\Delta\vec{X}=(\Delta p_m,\Delta
p_{\mathrm{diff,1}})$.  $\int {\cal L} dt$ represents the integrated
luminosity, $N(\Delta\vec{X})$ the number of true $(e,e'pp)$ events,
and ${\cal V}(\Delta\vec{X})$ the experimental detection volume in
phase space.  The factor $|\partial T_2/\partial E_x|$ is a Jacobian. 

The detection volume is calculated by a Monte-Carlo method using 
$10^8$ events within a nine-dimensional volume $\cal V$.  It
takes into account the energy and angular acceptances of the QDQ and
both HADRON detectors. The sampling error due to this integration adds
only 0.4\% to the statistical uncertainty of the calculated 
cross sections.

The integration over $E_x$ is performed in the range
$-11<E_x<14~\mbox{MeV}$.  The lower energy is set at a value
corresponding to $4\sigma$ of the peak width.  The events in the
region at $E_x>0~\text{MeV}$ correspond to true
$^{3}\mbox{He}(e,e'pp)$ events including events of which either the
incident or the scattered electron lost energy due to the emission of
a photon, resulting in a reconstructed $E_x$ that is systematically
larger than zero. The shape of this radiative tail as a function
of the excitation energy was calculated using the formalism
of~\cite{mot69}.  The upper integration limit was set at $E_x$=14~MeV. 
For each kinematic setting, the fraction of events beyond this cutoff
was calculated and applied as a correction factor to the data; this
factor varies from 1.14 to 1.16.  In selecting this region in excitation
energy, the uncertainty due to radiative effects on the other
kinematic quantities is negligible.

\begin{figure}
    \centerline{\epsffile{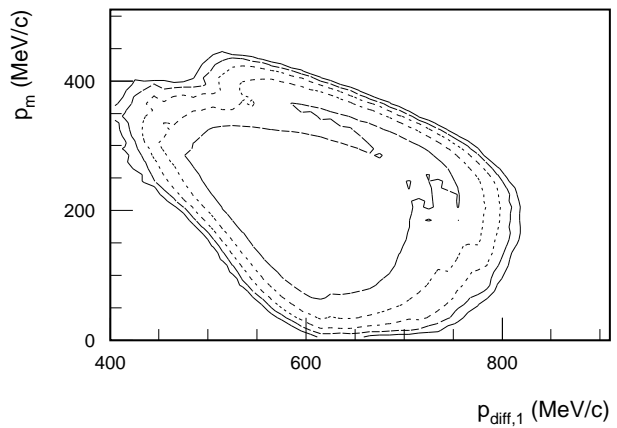}}
    \caption{
		Contour plot of the detection volume ${\tilde{\cal V}}$ as a
		function of $(p_m,p_{\mathrm{diff,1}})$ for the combination
		of the kinematic settings LQA, LQV, and PEF.  Contours are
		drawn at 30\%, 10\%, 3\%, 1\%, and 0.3\% of the maximum value.
		\label{fig:nicephaspa}
		\label{fig:nicedetvol}
	}
\end{figure}

Because of limitations imposed by the analysis software, the
integration over the $E_x$ interval is performed on the measured yield
and the detection volume separately, yielding an eight-fold
differential detection volume ${\tilde{\cal V}}$.  In
Fig.~\ref{fig:nicephaspa} the coverage of the $(p_m,p_{\mathrm{diff,1}})$
phase space by the detection volume ${\tilde{\cal V}}$ is shown for
the three overlapping kinematic settings at $q$=305~MeV/$c$.

With the experimental cross sections a systematic error of 8\% is
associated.  This systematic error is the quadratic sum of the
uncertainties in the target thickness as deduced from elastic
scattering measurements (3\%), the detector efficiency
simulations~(6\%), the dead time of the electronics of the HADRON
detectors~(2\%) and by other sources contributing less than 1\%
each~\cite{grothe}.

\subsection{Phase-space averaged theoretical cross section}

The comparison of the theoretical model~\cite{gol95} to the data
requires calculation of the cross section for specific kinematic
configurations, as well as the evaluation and averaging of the cross
section over the experimental detection volume.  The theoretical cross
section depends on seven kinematic variables that uniquely define the
configuration.  In general, the data are presented as a function of
two or three quantities, derived from the basic kinematic variables. 
In this way an implicit averaging over the other quantities within the
experimental detection volume is performed.

For a fair comparison between theory and data, the same averaging
should be applied to the calculated cross sections.  This averaging
cannot be performed analytically because of the complexity of the
integration limits, \emph{i.e.}, the shape of the experimental
detection volume.  Therefore, the integrals were approximated by a sum
over an orthogonal grid.  This averaging of the cross section over the
experimental detection volume of each kinematics was performed for the
central value of $(\omega,q)$ only, because of constraints on the
available computational resources.  This introduces a systematic
uncertainty, which is estimated to be less than 6\%, mainly due to the
dependence of the cross section on the transferred three-momentum.

For a given interval in the variables in which the cross
section is presented, \emph{e.g.}, an interval $\Delta p_m$, the
average cross section is defined as
\begin{equation}
  \left\langle\frac{d^8\sigma}{dV^8}\right\rangle
		\bigg\rfloor_{\Delta p_m} =
    \frac{
      \int \frac{d^8\sigma}{dV^8}(\vec{v})
           D(p_m(\vec{v});\Delta p_m) D(\vec{v};\vec{A}) d\vec{v}
    }{
      \int D(p_m(\vec{v});\Delta p_m) D(\vec{v};\vec{A}) d\vec{v}
    },
  \label{eq:theoavg}
\end{equation}
where $\vec{v}$ is the vector representing the quantities 
$(\theta_1,\phi_1,\theta_2,\phi_2,T_1)$ in the laboratory system, 
$\vec{A}$ the acceptance region of the experimental detection setup and
$D(\vec{x};\vec{R})$ a two-valued function that is only
different from zero if $\vec{x}$ is inside the region $\vec{R}$.

\begin{figure}
    \centerline{\epsffile{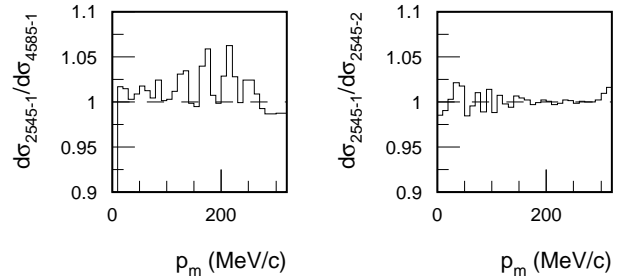}}
    \caption{
		Theoretical cross section for the LQ kinematic setting as a
		function of the neutron momentum, averaged over the
		experimental detection volume.  The panels
		show the ratio of the cross section calculated with the finest
		grid to two coarser grids.  The numeric labels in these two
		panels specify the grid spacing in degrees and in MeV:  
		$d\sigma_{\Delta \theta_1
		\Delta\phi_1 \Delta \theta_2 \Delta \phi_2 - \Delta T_1}$ 
		\label{fig:gridtest}
	}
\end{figure}

The two integrals are approximated by their sums, determined with
equidistant, orthogonal grids in the laboratory quantities $\vec{v}$. 
The distance between the gridpoints was chosen such that the
resulting error introduced in the final result is below 6\%.
This corresponds to $2.5\times 10^6$ cross-section calculations per
kinematic setting for every current operator used.  To
verify the accuracy obtained, the cross section was calculated with a
varying amount of grid points; the results are
displayed in Fig.~\ref{fig:gridtest}.  It was concluded from these and
similar tests that a grid point density of $(\Delta \theta_1, \Delta
\phi_1, \Delta \theta_2, \Delta \phi_2, \Delta T_1)=
(2^{\circ},5^{\circ},4^{\circ},5^{\circ},1~\text{MeV})$ is
sufficiently accurate.

The number of two-body angular momenta $j$ taken into account in the
$NN$ interaction also influences the calculated cross sections.  
Two-body angular
momenta in the $NN$ potential up to $j=3$ have been taken into account
to reach a result converged to approximately 6\%.

\section{Results}
\label{sec:res}

The cross section of the $^{3}\mbox{He}(e,e'pp)$ reaction depends on
seven independent kinematic variables.  However, the statistical
accuracy of the data does not allow representation of the measured
cross section for small intervals in all seven quantities
simultaneously.  The character of the hadronic current operator and
the $^{3}\mbox{He}$ bound-state wave function suggest that a limited
set of observables carries the characteristic information of the
$^{3}\mbox{He}(e,e'pp)$ process.

The momentum distributions for a nucleon pair in $^{3}\mbox{He}$ shown in
Fig.~\ref{fig:wavefunc}, suggest an important role for the relative
momentum ($\vec{p}_{\mathrm{rel}}$) and the pair
momentum~($\vec{p}_{\mathrm{cm}}$). 
Therefore, the missing momentum $p_m$, which in a
direct $(e,e'pp)$ reaction mechanism reflects the neutron momentum in
the initial state, is selected as an observable.  The electron
kinematics naturally define two relevant observables:  the energy
transfer $\omega$ and the momentum transfer $q$.  Alternatively, at
fixed $q$, the energy transfer $\omega$ can be replaced by the
invariant energy $W_{N'N'}$ of the two nucleons of a struck pair in the
final state. 

Another significant process that influences the cross section is
the rescattering among the outgoing nucleons. In particular, when two
nucleons are emitted with (vectorially) comparable momenta the
cross section will be notably enhanced. Within the experimental
detection volume, such `FSI configurations' occur between the
forward proton and the neutron. Hence, the momentum
difference of these two nucleons was selected as an observable:
\begin{equation}
        \vec{p}'_{ij} = \vec{p}'_i - \vec{p}'_j.
\end{equation}
The `FSI configuration' corresponds to $p'_{ij} \rightarrow
0~\mbox{MeV}/c$. 

Investigation of the coupling mechanisms by one-body currents shows a
dominant role for coupling of the virtual photon to the forward
proton.  In this case, the relative momentum $\vec{p}_{\it rel}$ of
two nucleons in the initial state can be related to the momentum
\begin{equation}
  \label{eq:pdiff}
  \vec{p}_{\mathrm{diff,1}} = 
    (\vec{p}'_1 - \vec{q}) - \vec{p}'_2 \triangleq \vec{p}_1 -
    \vec{p}_2 \equiv 2\vec{p}_{\mathrm{rel}}.
\end{equation}

In the following sections, the data are presented as a function of these
observables and compared to the results of continuum-Faddeev
calculations.

\subsection{Neutron momentum distribution}

\begin{figure}
    \centerline{\epsffile{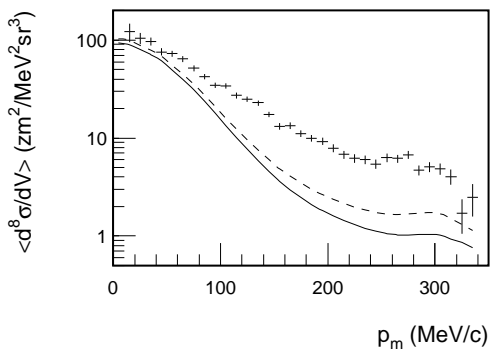}}
    \caption{
		Average cross section as a function of $p_m$ for the
		data measured in kinematic settings LQA and LQV. The solid
		and dashed lines represent the results of calculations 
		with a one-body current operator and including MECs,
		respectively, using the Bonn-B $NN$ potential. 
		\label{fig:pmg1slic}
		\label{fig:pmlq}
	}
\end{figure}

The differential cross sections measured at the kinematics LQ are shown
as a function of $p_m$ in Fig.~\ref{fig:pmlq}. They are averaged over the
detection volume corresponding to the settings LQA and LQV,
\emph{i.e.}, $40^{\circ}<\theta_{H3}<68^{\circ}$ and
$-140^{\circ}<\theta_{H4}<-72^{\circ}$.  The cross section decreases
roughly exponentially as a function of the neutron momentum between
zero and 300~MeV/$c$.  This reflects the neutron momentum distribution
inside $^{3}\mbox{He}$ for relative momenta in the $pp$ pair between
250~and 330~MeV/$c$ per nucleon, the region probed in this kinematic
configuration~(\emph{cf.} Fig.~\ref{fig:nicedetvol}). 

Signatures of two-proton knockout by one-body hadronic currents will
most likely be found at low $p_m$. In this domain the neutron is left
with a small momentum and can be considered as a spectator, since in
direct $pp$ knockout contributions from two-body currents are
suppressed.  As mentioned in section~\ref{sec:theory}, 
contributions from MECs are 
prohibited in a non-relativistic framework, as the photon will 
not couple to the neutral mesons
exchanged in the $pp$ pair.  Additionally, the knockout via $pp
\rightarrow \Delta^{+}p \rightarrow pp$ is suppressed since the
otherwise dominant M1 transition is forbidden by angular momentum and
parity conservation for protons initially in a $^1S_0$
state~\cite{wil96}.

A comparison for $p_m \lesssim 100~\mbox{MeV}/c$ with the results of
continuum Faddeev calculations including only one-body currents shows
a fair agreement; they account for approximately 50 to 80\% of the
measured strength in this region, while the contribution of MECs is
small (5\%).  At higher missing-momentum values, one-body calculations
underestimate the data by a factor of five.  The high missing-momentum
region is likely to be dominated by two-body hadronic currents~(MECs and
ICs), which involve coupling of the virtual photon to a proton-neutron
pair.  Such processes predominantly contribute to the
$^{3}\mbox{He}(e,e'pp)$ cross section at large $p_m$, because in such a
process the emitted neutron has a large energy.  This
expectation is supported by the results of calculations with MEC
contributions, which show an increased importance of MECs of up to
40\% of the calculated strength at $p_m\approx 300~\mbox{MeV}/c$, as
compared to the low $p_m$ region. 

In the high $p_m$ domain also a sizeable contribution from 
$\Delta$-excitation can be expected:  excitation of a $\Delta$ within a $pn$
pair, a process that is not suppressed by selection rules like in the
$pp$ case, will contribute to the cross section primarily in this domain. 

\begin{figure}
    \centerline{\epsffile{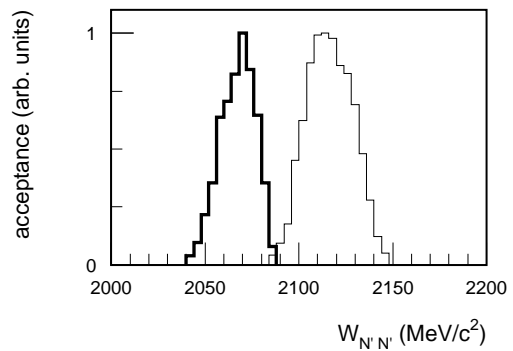}}
    \caption{
		Invariant mass of the two nucleons in the final state for
		the $p'_1p'_2$ pair at low $p_m$ (thick line) and 
		the $p'_1n'$ pair at high $p_m$ (thin line) in 
		the kinematics LQ. 
		\label{fig:wppnlq}
	}
\end{figure}

Excitation of the $\Delta$ resonance strongly depends on the invariant
mass of the $\gamma^{*}NN$ system.  If one considers a direct reaction
on a proton pair at small $p_m$ values, the invariant mass $W_{\gamma
pp}$ in the initial state can be identified with the final-state
observable $W_{p'_1p'_2}$.  At LQ this invariant mass ranges from 2050
to 2080~MeV/$c^2$, which is well below the mass of the $\Delta N$
system.  If one assumes absorption of the virtual photon on a $pn$
pair, the relevant invariant energy is that of the $\gamma^{*}pn$
system.  The corresponding invariant mass $W_{p'_1n'}$ ranges from
2100 to 2140~MeV/$c^2$ for $p_m$
values around 300~MeV/$c$~(see Fig.~\ref{fig:wppnlq}).  The invariant
mass of the other $pn$ pair, \emph{i.e.}, $W_{p'_2n'}$, is similar to
that of the $pp$ pair for this $p_m$ region.  Therefore the cross
section for intermediate $\Delta$ excitation in the $p_1n$ pair will
be dominant.

Calculations of the $^{16}\text{O}(\gamma,pn)$ cross section indicate
a strong dependence of the contribution of isobar currents on the
photon energy~\cite{mac93}.  These calculations, as well as
calculations of photon-induced deuteron breakup, which use a different
$\Delta$ propagator~\cite{wilb96}, indicate a maximum in the cross
section due to $\Delta$ excitation around $E_\gamma \sim
250~\text{MeV}$. This corresponds to an invariant mass $W_{p'_1n'}$ 
around 2125~MeV/$c^2$, which is at the center of our acceptance for
the LQ kinematic setting.

\begin{figure}
    \centerline{\epsffile{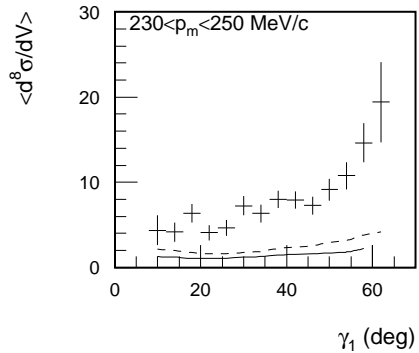}}
    \caption{
		Average cross section as a function of $\gamma_1$ for the
		$p_m$ interval from 230 to 250~MeV/$c$ for the combined
		kinematic settings at $q$=305~MeV/$c$. Curves as in
		Fig.~\protect\ref{fig:pmlq}.
		\label{fig:gamma1}
	}
\end{figure}

The excitation of the $\Delta$ resonance and the subsequent decay of
the $\Delta N$ system in a $pn$ pair is expected to cause --~due to
its multipole character~-- a characteristic angular dependence of the
cross section. This can also be seen in calculations of the
$^{16}O(\gamma,pn)$ cross section at $E_\gamma=281~\text{MeV}$, a
reaction that is also dominated by the isobar current~\cite{ryck94}. 
In this reaction the cross section reaches a maximum between
approximately $\gamma_1$=40$^{\circ}$ and 100$^{\circ}$, depending on
the proton energy $T_1$.  In Fig.~\ref{fig:gamma1} the angular
dependence of the measured $^{3}\mbox{He}(e,e'pp)$ cross sections is
shown for the missing momentum interval from 230 to 250~MeV/$c$.  The
data exhibit a characteristic angular dependence:  at larger angles
$\gamma_1$ a strong increase is observed that is not reproduced by the
calculation based on a one-body current operator or those including
MECs.

\subsection{Momentum-transfer dependence\label{sec:qdepres}}

Further information on the reaction mechanism is obtained from the
dependence of the cross section on $q$.  The data at the various $(\omega,q)$
points all show a similar dependence of the cross section on
the missing momentum.  This is expected for quasi-free two-proton
knockout, in which the neutron acts as a spectator, implying that the
effects due to final-state interactions are generally small.  However,
strong rescattering effects may occur at specific values of $p_m$,
because $p'_{13} \rightarrow 0~\mbox{MeV}/c$, the exact position of which
depends on the experimental detection volume.  Therefore, no reliable
comparison can be made between data of different $(\omega,q)$
settings for kinematic domains in which an `FSI configuration' occurs. 
For the $q$-scan data, the rescattering effects limit the usable domain
to $p_m \lesssim 220~\mbox{MeV}/c$.

\begin{figure}
    \centerline{\epsffile{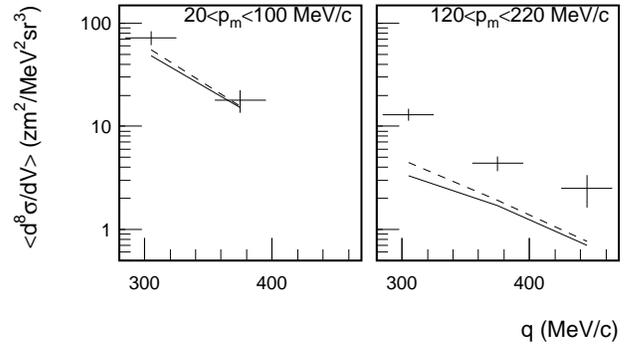}}
    \caption{
		Average cross-section dependence on the momentum transfer $q$,
		for two slices in the final-state neutron momentum at
		$\omega$=220~MeV. Curves are the same as in 
		Fig.~\protect\ref{fig:pmlq}.
		The data were averaged over the
		domain $10^{\circ}<\gamma_1<25^{\circ}$. The horizontal
		error bars indicate the range in $q$ values covered
		due to the acceptance of the spectrometer.  The domain
		$p_m<120~\mbox{MeV}/c$ is not covered by the detection
		volume of the HQ kinematic setting. 
		\label{fig:qdep}
	}
\end{figure}

In Fig.~\ref{fig:qdep} the cross section is shown as a function of $q$
for two slices in $p_m$.  The data at missing-momentum values below
100~MeV/$c$ show a decrease by a factor of four between
$q$=305~MeV/$c$ and $q$=375~MeV/$c$.  Both this slope and the absolute
magnitude of the cross sections are reasonably well described by the
calculations.  For both values of the momentum transfer a calculation
with only one-body hadronic currents accounts for 72$\pm$13\% of the
measured strength.  The inclusion of MEC contributions has --~as
expected~-- only a minor effect and increases the calculated strength
to 80\% of the experimentally observed value.  The fair agreement
between data and theory for both momentum transfer values indicates
that, in the $p_m$ domain below 100~MeV/$c$, the cross section is
predominantly driven by a one-body reaction mechanism.

In the $p_m$ domain $120<p_m<220~\mbox{MeV}/c$, the difference between
a one-body calculation and data is about a factor of five.  Inclusion
of MECs in the calculation increases the calculated cross section by
10 to 35\%, depending on the momentum transfer, thus reducing the
discrepancy to about a factor of four.  The $q$-dependence of the data
and the calculations is nevertheless similar.

\subsection{Energy-transfer dependence}

As discussed in the previous section, the low $p_m$ region is most
likely due to direct two-proton knockout, as in this domain the
neutron is left `at rest'.  In case of such a direct reaction
mechanism, the invariant mass of the two emitted protons
$W_{p'_1p'_2}$ can be identified with the invariant mass of the
$\gamma pp$ system.  For
$p_m<100~\mbox{MeV}/c$, this invariant mass ranges from 2055~MeV/$c^2$
at $\omega=220~\mbox{MeV}$~(well below the $\Delta$ resonance) to
2120~MeV/$c^2$ at $\omega=290~\mbox{MeV}$, \emph{i.e.}, almost on top
of the resonance.

\begin{figure}
    \centerline{\epsffile{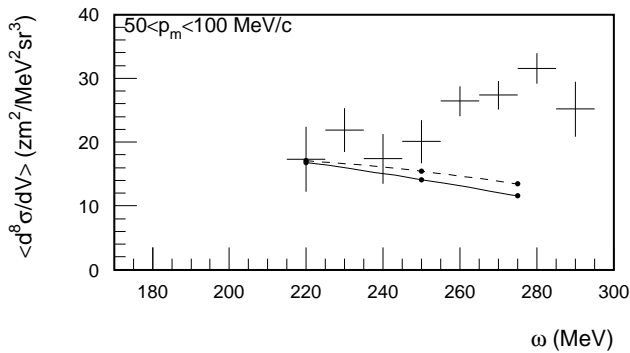}}
    \caption{
		Average cross section as a function of the energy transfer
		$\omega$ at $q$=375~MeV/$c$ and $50<p_m<100~\mbox{MeV}/c$. 
		Curves as in Fig.~\protect\ref{fig:pmlq}.
		\label{fig:omegalow}
	}
\end{figure}

In Fig.~\ref{fig:omegalow} the data for the $p_m$ domain from 50 to
100~MeV/$c$ at $q$=375~MeV/$c$ are displayed as a function of the
energy transfer $\omega$.  The calculations performed with a one-body
current operator show a slightly decreasing trend as a function of
$\omega$, which is due to changes in the relative
momentum of the $pp$ pair in the initial state probed in the reaction;
whereas at $\omega$=220~MeV the central value for the relative
momentum is 290~MeV/$c$ per nucleon, it has risen to 360~MeV/$c$ per
nucleon at $\omega$=275~MeV.

As expected from the data shown in section~\ref{sec:qdepres},
the agreement between data and calculations for
$\omega\approx 220~\mbox{MeV}$ is quite good, which can be considered as
evidence for the dominance of one-body currents in this $p_m$ and
$\omega$ domain.  The inclusion of MECs in the calculation hardly
changes the cross section at $\omega$=220~MeV. 

At energy transfer values in the range from 220 to 290~MeV, the data
show an increase of almost 50\%.  The contribution due to MECs remains
rather low~(below 15\%).  Therefore, the increase of the experimental
cross section probably reflects an increasing importance of
intermediate $\Delta$ excitation at higher invariant masses
$W_{p'_1p'_2}$.

\begin{figure}
    \centerline{\epsffile{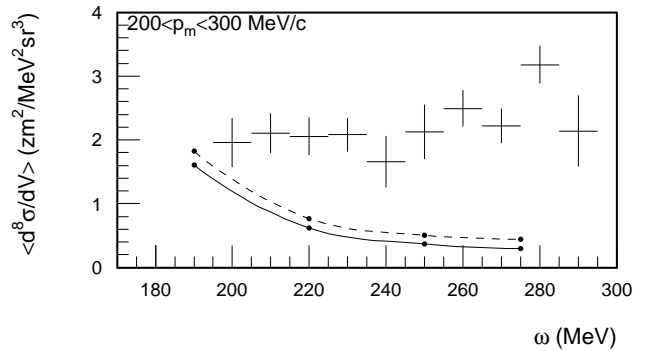}}
    \caption{
		Average cross section as a function of the energy transfer
		$\omega$ at $q$=375~MeV/$c$ and $200<p_m<300~\mbox{MeV}/c$.
		Curves as in Fig.~\protect\ref{fig:pmlq}.
		\label{fig:omegahi}
	}
\end{figure}

Similarly, the cross section as a function of $\omega$ for
the missing momentum region from 200 to 300~MeV/$c$ is shown in
Fig.~\ref{fig:omegahi}.  The cross sections calculated with
a one-body current operator decrease systematically for increasing
values of $\omega$, again due to the increasing relative momentum of
the protons in the $pp$ pair.  In addition, at
$\omega~\approx~200~\mbox{MeV}$, the kinematical variables are close
to an `FSI configuration' occuring within the experimental detection
volume at $p_m=320~\mbox{MeV}/c$.  The measured cross section does not
show pronounced dependence on $\omega$.  At the lowest $\omega$ value,
the ratio of the experimental and theoretical cross section is
1.6$\pm$0.3, whereas at higher values of the energy transfer the data
overshoot the theoretical results by about a factor of five. 

In the $p_m$ domain probed here, a considerable part of the strength
may be due to coupling of the virtual photon to a $pn$ pair, whose
invariant mass in the final state is considerably larger than in
the $\gamma pp$ system:  $2110<W_{p'_1n'}<2190~\mbox{MeV}/c^2$ for
$250<p_m<300~\mbox{MeV}/c$.  This domain corresponds to the region
where the total cross section for photon-induced deuteron breakup
reaches its maximum, \emph{i.e.}, at $E_\gamma \approx
265~\mbox{MeV}$~\cite{wil96}.  This corresponds to an invariant mass
of the $\gamma pn$ system of approximately 2140~MeV/$c^2$.  In this
energy range, the photo-induced breakup of the deuteron is known to be
dominated by intermediate $\Delta$ excitation and its subsequent
decay.

\subsection{Rescattering configurations}

\begin{figure}
    \centerline{\epsffile{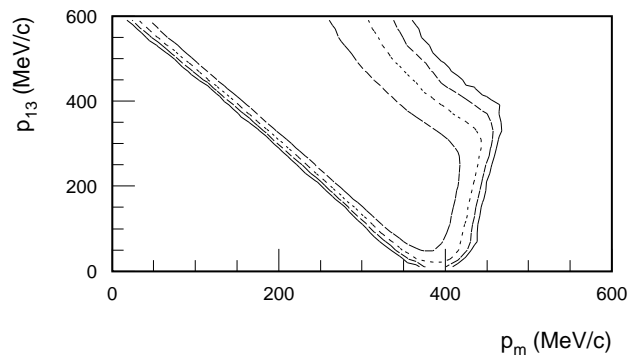}}
    \caption{
		Projection of the detection volume on the $(p'_{13},p_m)$ plane
		for kinematics HW. 
		\label{fig:phasp13hw}
	}
\end{figure}

For kinematic domains in which two nucleons are emitted with similar
momentum vectors, rescattering effects can influence the cross section
considerably.  The presentation of the data as a function of the momentum
difference $p'_{ij}$ allows an investigation of these rescattering
effects.

In the HQ kinematic setting the detection volume extends to
$p'_{13}=0~\mbox{MeV}/c$.  A good agreement between the data obtained
at $\omega=220~\mbox{MeV}$ at $p_m$ values below 100~MeV/$c$ and
continuum-Faddeev calculations based on a one-body hadronic current
was shown already in Ref.~\cite{groprl}.  Similar configurations occur
at other kinematic settings.  As the cross section depends strongly on
both $p'_{13}$ and $p_m$ individually and the detection volume is
non-rectangular in these two observables (see
Fig.~\ref{fig:phasp13hw}), one can reduce these rescattering effects
by limiting the $p_m$ range.

\begin{figure}
    \centerline{\epsffile{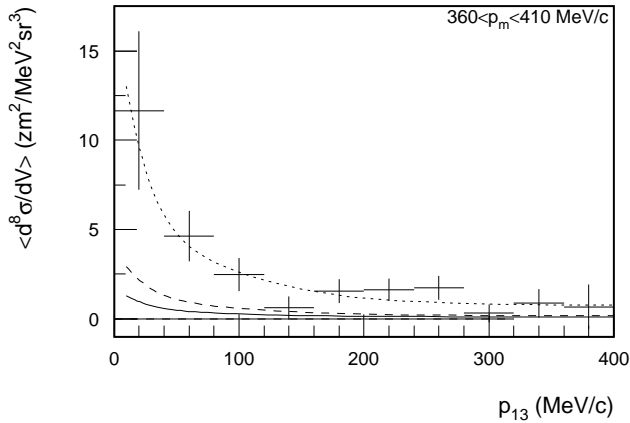}}
    \caption{
		Average cross section for the `FSI configuration' at 
		kinematics HW.  The $p_m$ acceptance has been limited to
		$360<p_m<410~\mbox{MeV}/c$.  The curves are results of the
		Faddeev calculations with a one-body current operator (solid),
		including MECs (dashed) and a scaled ($\times 4.43$) MEC result
		(dotted), all based on the Bonn-B potential.
		\label{fig:p13hw}
	}
\end{figure}

In particular the HW kinematics contains a fairly broad region in $p_m$
--~between 360 and 410~MeV/$c$~-- for which the region around 
$p'_{13}=0~\mbox{MeV}/c$ is covered by the detection volume~(see
Fig.~\ref{fig:phasp13hw}).  Unfortunately, the high value of energy
transfer together with the high $p_m$ region means that a sizeable
part of the reaction occurs via intermediate $\Delta$ excitation in
the $pn$ pair. This has the consequence that the calculated cross section,
even including MECs, globally
underestimates the data by a factor of 4.4 at
$360<p_m<410~\mbox{MeV}/c$ (and by a factor of 9.1 with respect to a
one-body calculation) as can be seen from Fig.~\ref{fig:p13hw}.

Although the absolute magnitude is not correctly predicted, the
dependence of the cross section on $p'_{13}$ is well reproduced by
both the one-body calculations and those including MECs.  The
similarity in shape between both types of calculations suggests that
the dependence of the cross section on $p'_{13}$ is mainly due to $NN$
rescattering and the magnitude to the current operators used. 
Scaling of the calculated cross sections that include MECs, by a
factor of 4.43 results in a good agreement between data and
calculations over the entire $p'_{13}$ domain.  Hence, although the
calculations do not adequately describe the absorption of the virtual
photon by the nucleon pairs, it is likely that the continuum Faddeev
calculations adequately describe final-state rescattering effects.

\subsection{Relative momenta and potential models}

An investigation of the data at low missing momentum, \emph{i.e.},
$p_m \lesssim 100~\mbox{MeV}/c$, and at an energy transfer value of
$\omega$=220~MeV, showed a dominant contribution from direct
two-proton knockout by a one-body hadronic current.  As argued,
breakup induced by coupling to a one-body current in principle allows
investigation of the $^{3}\mbox{He}$ bound-state wave function.  The
calculations indicate that the cross section in this domain is almost
exclusively determined by coupling of the virtual photon to the
forward proton.  Hence, according to the calculations, the observable
$p_{\mathrm{diff,1}}$~as defined in Eq.~(\ref{eq:pdiff}) should be
representative of the initial-state proton momentum $p_1$. 
Investigation of the cross section as a function of $p_{\mathrm{diff,1}}$
in the low $p_m$ domain at LQ may thus lead to insight in the
initial-state wave function of $^{3}\mbox{He}$.

In Fig.~\ref{fig:wavefunc}, the probabilities associated with the
$^{3}\mbox{He}$ wave function were shown as a function of the relative
momentum of a nucleon pair for various $NN$ potentials. 
The shape of the wave functions within the experimentally accessible
domain is similar for Bonn-B, CD-Bonn and Nijmegen-93.  The result for
the Argonne $v_{18}$ potential is different, especially in the high
$p_{\mathrm{cm}}$ and high $p_{\mathrm{rel}}$ region, but as in this domain 
two-body currents are expected to give a significant contribution to
the $^{3}\mbox{He}(e,e'pp)$ cross section,
no quantitative comparison to the data can be
made.  However, small differences in magnitude exist among the
various model predictions for low $p_m$, and also here the difference
is largest for Argonne $v_{18}$.  One expects for all potential models
a decrease of the experimental cross section due to one-body currents
as a function of $p_{\mathrm{diff,1}}$, because the probability density 
decreases as a function of $p_{\it rel}$.  At higher missing
momenta, the cross section dependence becomes increasingly more flat.

\begin{figure}
    \centerline{\epsffile{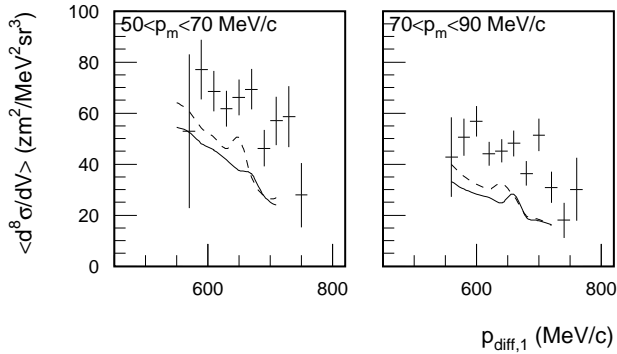}}
    \caption{
		Average cross section as a function of 
		$p_{\mathrm{diff,1}}$ for two slices in $p_m$ of 
		20~MeV/$c$ wide.  Data are taken
		from the combined kinematic settings LQA, LQV and PEF.
		Solid curves are based on one-body currents only; 
		dashed curves include MECs, both calculated using 
		the Bonn-B potential.  The
		wiggles in the calculated cross section are due to small
		variations in the parts of the detection volume that
		contribute in the different kinematic configurations.
		\label{fig:pdiffpm}
	}
\end{figure}

The rapid changes in cross section as a function of $p_m$ make it
necessary to investigate the dependence on $p_{\mathrm{diff,1}}$ for slices in
the missing momentum, that are not wider than 20~MeV/$c$.
The data in the two graphs of Fig.~\ref{fig:pdiffpm} are taken 
from adjacent slices in $p_m$ (from 50--70 and from 70--90~MeV/$c$).
These already show a different  dependence on $p_{\it diff}$.  
The fine binning thus
required leads to a reduced statistical accuracy for the measured
cross sections.

It was already shown in Fig.~\ref{fig:pmlq} and Fig.~\ref{fig:qdep}
that in the domain $p_m\lesssim 100~\mbox{MeV}/c$ the dependence of
the cross section on $p_m$ and $q$ is fairly well reproduced by
calculations performed with the Bonn-B potential. 
Figure~\ref{fig:pdiffpm} shows that also the dependence of the cross
section on the momentum difference $p_{\mathrm{diff,1}}$
(\emph{c.f.}~Eq.~\ref{eq:pdiff}) in this $p_m$ domain is quite well
reproduced by the calculations. 

\begin{figure}
    \centerline{\epsffile{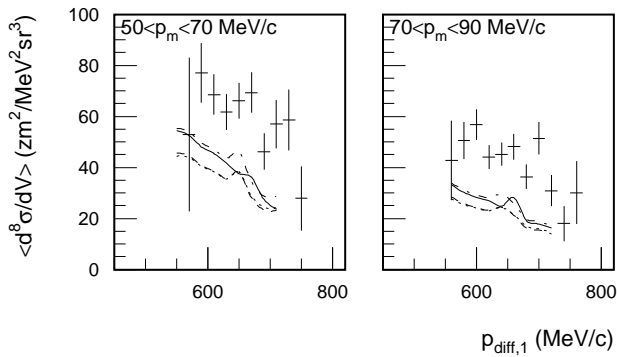}}
    \caption{
		Average cross section as a function of $p_{\mathrm{diff,1}}$ as
		in Fig.~\protect\ref{fig:pdiffpm}.  All curves are based on
		one-body currents only, but calculated using various models
		for the $NN$ potential:  solid curves Bonn-B, dashed curves
		CD-Bonn, dotted curves  Argonne $v_{18}$, dot-dashed curves
		Nijmegen-93.
		\label{fig:pdiffpots}
	}
\end{figure}

In Figure~\ref{fig:pdiffpots} the same data are compared to
predictions from continuum Faddeev calculations, performed with
different $NN$-potential models.  Differences in both
magnitude and slope are observed, with the Argonne $v_{18}$ prediction
being up to 15\% lower than the one based on Bonn-B. 

The variations between the calculations performed with the various
models are of the same order of magnitude as the effects of MECs,
which was only calculated using the Bonn-B potential.  The influence
of intermediate $\Delta$ excitation on the calculated slope is as of
yet unknown; also the underestimation of the data by all four
calculations, which amounts to approximately 30\% at
$50<p_m<90~\mbox{MeV}/c$, is still not explained quantitatively.  In
the missing-momentum region above 200~MeV/$c$, the differences in the
calculated cross section due to the $NN$ potential are almost
negligible within the experimentally probed domain.  In view of these
uncertainties, the low $p_m$ data do not yet suggest a preference for
any of the potential models.

\section{Summary and Conclusions}

In summary, the cross sections for the exclusive reaction
$^{3}\text{He}(e,e'pp)n$ were measured for three values of the
three-momentum transfer of the virtual photon~($q$=305, 375, and
445~MeV/$c$) at an energy transfer value $\omega$ of 220~MeV.  At
$q$=375~MeV/$c$, measurements were performed over a continuous range
of transferred energy from 170 to 290~MeV.  At $q$=305~MeV/$c$, a
large range in center-of-mass ($p_{\mathrm{cm}}$=0 to 310~MeV/$c$) and
$pp$ relative ($p_{\mathrm{diff,1}}$=500 to 800~MeV/$c$) momenta was
covered.  The data are compared to results of continuum Faddeev
calculations that account for the contributions of rescattering among
the emitted nucleons.  These calculations include both one-body
hadronic currents as well as contributions due to the coupling to
$\pi$ and $\rho$ mesons in an intermediate state.  Various potential
models were used in the calculations:  Bonn-B, charge-dependent (CD)
Bonn, Nijmegen-93 and Argonne~$v_{18}$. 

Calculations performed with only a one-body hadronic current operator
show a fair agreement with the data obtained at $p_m \lesssim
100~\text{MeV}/c$ at $\omega=220~\text{MeV}$ and $q$=305~MeV/$c$. 
Measurements performed at $q$=375~MeV/$c$ show similar results.  Here,
the inclusion of MECs in the current operator only has a minor effect on
the calculated strength.  It can therefore be concluded that at
$\omega$=220~MeV and $p_m<100~\text{MeV}/c$ the cross section is
dominated by direct knockout of two protons via a one-body hadronic
current.  At higher $p_m$ values, from 120 to 320~MeV/$c$, a
discrepancy of up to a factor of five is observed between the data and
calculations with a one-body current operator only.  Contributions due
to MECs increase the calculated strength by up to 35\% at most.

The influence of intermediate $\Delta$ excitation depends on the
invariant mass of the $\gamma^{*}NN$ system involved.  To investigate
such isobar currents, measurements were performed in the domain
$\omega$=170--290~MeV.  This range corresponds for
$p_m<100~\text{MeV}/c$ to invariant masses $W_{p'_1p'_2}$ between 2055
and 2120~MeV/$c^2$.  An increase of the measured cross section by
almost 50\% is seen over this range in energy transfer. 
Theoretical predictions including MECs underestimate the data from
30\% at $\omega$=220~MeV to a factor of two at the higher $\omega$
values, presumably reflecting the increased importance of the $\Delta$
resonance. 

At higher neutron momentum values, data and theoretical predictions
differ up to a factor of five for all values of $\omega$.  This is
likely due to intermediate $\Delta$ excitation by the virtual photon
of the $pn$ pair, for which the invariant mass in the final state
amounts to approximately 2150~MeV/$c^2$. This value corresponds to the
position of the resonance in deuteron electrodisintegration.  A
further indication for the importance of intermediate $\Delta$
excitation as a process contributing to the $^{3}\text{He}(e,e'pp)$
cross section in the $p_m$ domain above 100~MeV/$c$ can be found in the
dependence of the cross section on the forward proton emission angle
$\gamma_1$.  Comprehensive treatment of the $\Delta$ degrees-of-freedom
within the continuum Faddeev framework is necessary before
quantitative conclusions can be drawn from the data measured at high
$p_m$ or high $\omega$ values.

Within the experimental detection volumes covered in the measurement
at $(\omega,q)=(220~\mbox{MeV}, 445~\mbox{MeV}/c)$ and
$(275~\mbox{MeV}, 375~\mbox{MeV}/c)$ and a selected domain in $p_m$,
two nucleons are emitted with a low relative momentum in the final
state.  In these regions, rescattering effects strongly influence the
cross section.  Data from such specific `FSI configurations' provide a
good tool to check the calculations in this respect.  Good agreement
was found between data measured at $\omega$=220~MeV and
$q$=445~MeV/$c$, and theoretical predictions, when presented as a
function of the $pn$ momentum difference in the final state.  Data
obtained at $\omega$=275~MeV and $q$=375~MeV/$c$ confirmed this
result, although the absolute magnitude of the cross section is
underestimated by the predictions, probably due to lack of isobar
contributions in the current operator used.

Information on the relative momentum of the two protons in $^{3}\text{He}$
may be obtained in
the domain $\omega \approx 220~\text{MeV}$ and $p_m<100~\text{MeV}/c$,
since in this domain knockout is expected to be dominated by direct
$pp$ emission via a one-body hadronic current.  The observed decrease
of the cross section as a function of relative momentum reflects the
behaviour of the wave function and is well reproduced by calculations
at low $p_m$.  Calculations performed with different models of the
$NN$ interaction lead to different predictions of the cross section
in this domain, both in magnitude and as a function of $p_{\text{rel}}$. 
The statistical and systematic uncertainties of the data, as well as the
sizeable changes induced in the predictions by the MECs and the
unknown influence of isobar contributions, do not yet permit to
express preference for any of the potential models considered. 

Larger differences between the wave functions calculated from the
various $NN$ potentials are observed at high center-of-mass momentum
values and for relative momenta above 400~MeV/$c$ per nucleon. 
However, the interpretation of this domain awaits either a better
theoretical treatment of the high $p_m$ region or experimental means
to isolate the contribution of isobar currents to the cross section. 
In this respect, separation of the $^{3}\text{He}(e,e'pp)$ cross
section in its contributing structure functions and an investigation
of the complementary reaction $^{3}\text{He}(e,e'pn)$ will provide
valuable information for better understanding the processes involved.

\section{Acknowledgements}

This work is part of the research program of the Foundation for
Fundamental Research on Matter (FOM) and was sponsored by the
Stichting Nationale Computerfaciliteiten (National Computing
Facilities Foundation, NCF) for the use of supercomputer facilities. 
Both organisations are financially supported by the Netherlands
Organisation for Scientific Research (NWO).  The support of the
Science and Technology Cooperation Germany-Poland, the Polish
Committee for Scientific Research (grant No.  2P03B03914), and the US
Department of Energy is gratefully acknowledged.  Part of the
calculations have been performed on the Cray T90 and T3E of the John
von Neumann Institute for Computing, J\"ulich, Germany.


\begin{references}
\bibitem[*]{co}
	Electronic address: Eddy.Jans@nikhef.nl.
\bibitem[\dagger]{goaddr}
	Present address: University of Illinois at Urbana-Champaign,
		1110 West Green Street, Urbana, IL 61801-3080.

\bibitem{glo96}
	W.~Gl\"ockle, H.~Wita\l{}a, D.~H\"uber, H.~Kamada and J.~Golak,
	Phys. Rep.\ {\bf 274}, 107 (1996).
\bibitem{car98}
	J.~Carlson and R.~Schiavilla, Rev. Mod. Phys. {\bf 70}, 743
	(1998).
\bibitem{Nogga}
	A.~Nogga, H.~Kamada and W.~Gl\"ockle, Phys. Rev. Lett. {\bf 85}, 944
	(2000).
\bibitem{glo99}
	W.~Gl\"ockle, H.~Kamada, J.~Golak, H.~Wita\l{}a, S.~Ishikawa and
	D.~H\"uber, in \emph{Proceedings of the Second Workshop on
	Electronuclear Physics with Internal Targets and the BLAST detector},
	eds. R. Alarcon and R. Milner, World Scientific (1999), pp. 185-200
	and references therein.
\bibitem{aud89}
	G.~Audit \emph{et al.}, Phys. Lett. {\bf B227}, 331 (1989).
\bibitem{aud91}
	G.~Audit \emph{et al.}, Phys. Rev. C.\ {\bf 44}, R575 (1991).
\bibitem{lag85}
	J.M.~Laget, Nucl. Phys. {\bf A446}, 489c (1985).
\bibitem{sar93}
	A.J.~Sarty \emph{et al.}, Phys. Rev. C\ {\bf 47}, 459 (1993).
\bibitem{aud97}
	G.~Audit \emph{et al.}, Nucl. Phys.\ {\bf A614}, 461 (1997).
\bibitem{kol96}
	N.R.~Kolb \emph{et al.}, Phys. Rev. C\ {\bf 54}, 2175 (1996).
\bibitem{emu94}
	T.~Emura \emph{et al.}, Phys. Rev. Lett.\ {\bf 73}, 404 (1994).
\bibitem{emu94b}
	T.~Emura \emph{et al.}, Phys. Rev. C\ {\bf 49}, R597 (1994).
\bibitem{zon95}
	A.~Zondervan \emph{et al.}, Nucl. Phys.\ {\bf A587}, 697 (1995).
\bibitem{kes95}
	L.J.H.M.~Kester \emph{et al.}, Phys. Rev. Lett.\ {\bf 74}, 1712
	(1995).
\bibitem{ond97}
	C.J.G.~Onderwater \emph{et al.}, 
	Phys. Rev. Lett.\ {\bf 78}, 4893 (1997).
\bibitem{ond98art}
	C.J.G.~Onderwater \emph{et al.}, 
	Phys. Rev. Lett.\ {\bf 81}, 2213 (1998).
\bibitem{ros97}
	G.~Rosner, in \emph{Proc.  Conf.  on Perspectives in
	Hadronic Physics}, Trieste 1997, ICTP/World Scientific, p.~185.
\bibitem{sta99art}
	R.~Starink \emph{et al.}, Phys. Lett. {\bf B474}, 33 (2000).
\bibitem{groprl}
	D.L.~Groep \emph{et al.}, Phys. Rev. Lett. {\bf 83}, 5443 (1999).
\bibitem{giu91}
	C.~Giusti and F.D.~Pacati, Nucl.Phys.\ {\bf A535}, 573 (1991).
\bibitem{wil96}
	P.~Wilhelm, J.A.~Niskanen and H.~Arenh\"ovel,
	Nucl. Phys.\ {\bf A597}, 613 (1996);
	Phys. Rev. Lett.\ {\bf 74}, 1034 (1995).
\bibitem{bof96} 
	S.~Boffi, C.~Giusti, F.D.~Pacati and M.~Radici, 
	\emph{Electromagnetic Response
	of Atomic Nuclei} (Clarendon Press, Oxford, 1996)
\bibitem{mei86}
	E.~van~Meijgaard and J.A.~Tjon, Phys. Rev. Lett.\ {\bf 57}, 3011
	(1986).
\bibitem{ish94}
	S. Ishikawa, J.~Golak, H.~Wita\l{}a, H. Kamada, W. Gl\"ockle, and
	D. H\"uber, Phys. Rev. {\bf C57}, 29 (1998) and references therein.
\bibitem{poo99}
	H.R.~Poolman, \emph{Quasifree Spin-dependent
	Electron Scattering from a Polarized ${^3}$He Internal Target},
	Ph.D.~thesis Vrije Universiteit Amsterdam 1999.
\bibitem{ank98}
	H.~Anklin \emph{et al.}, Nucl. Phys. {\bf A636}, 189 (1998).
\bibitem{spa98}
	C.M.~Spaltro \emph{et al.}, Phys. Rev. Lett. {\bf 81}, 2870 (1998).
\bibitem{gol95}
	J.~Golak, H.~Kamada, H.~Wita\l{}a, W.~Gl\"ockle and S.~Ishikawa,
	Phys. Rev. C\ {\bf 51}, 1638 (1995).
\bibitem{sch89}
	R.~Schiavilla, V.R.~Pandharipande and D.O.~Riska, 
	Phys. Rev. C\ {\bf 40}, 2294 (1989).
\bibitem{ris85}
	D. O. Riska, Phys. Scr. {\bf 31}, 471 (1985).
\bibitem{kot99}
	V. V. Kotlyar, H. Kamada, J.~Golak and W. Gl\"ockle, 
	Few-Body Systems {\bf 28}, 35 (2000).
\bibitem{chm99}
	K.~Chmielewski, S.~Nemoto, A.C.~Fonseca, P.U.~Sauer, Few-Body
	System Suppl. {\bf 10}, 335 (1999).
\bibitem{mac93}
	L.~Machenil, M.~Vanderhaeghen, J.~Ryckebusch and M.~Waroquier, 
	Phys. Lett. {\bf B316}, 17 (1993).
\bibitem{nog97}
	A.~Nogga, D.~Hueber, H.~Kamada, W.~Gl\"ockle, Phys. Lett. 
	{\bf B409}, 19 (1997).
\bibitem{vri90}
	L.~de Vries \emph{et al.}, 
	Nucl. Instr. Meth. in Phys. Res. {\bf A292}, 629 (1990).
\bibitem{pel99}
	A.R.~Pellegrino \emph{et al.}, 
	Nucl. Instr. Meth. in Phys. Res. {\bf A437}, 188 (1999).
\bibitem{mot69}
	L.W.~Mo and Y.S.~Tsai, Rev. Mod. Phys.\ {\bf 41}, 205 (1969).
\bibitem{grothe}
	D.L.~Groep, \emph{Correlations and Currents in $^3\mbox{He}$
	Studied with the $(e,e'pp)$ Reaction}, Ph.D.~thesis 
	Universiteit Utrecht 2000.
\bibitem{wilb96}
	Th.~Wilbois, P.Wilhelm, H.~Arenh\"ovel,
	Phys. Rev. C\ {\bf 54}, 3311 (1996).
\bibitem{ryck94}
	J.~Ryckebusch, L.~Machenil, M.~Vanderhaeghen, V.~van~der~Sluys and
	M.~Waroquier, Phys. Rev. C\ {\bf 49}, 2704 (1994).


\end{references}
\end{document}